\documentclass[aip,amsmath,amssymb,reprint,citeautoscript,groupedaddress,longbibliography]{revtex4-1}
\pdfoutput=1

\usepackage{graphicx}
\usepackage{siunitx}
\usepackage{natbib}

\newcommand{\dif}{\mathrm{d}}	

\newcommand{\mean}[1]{\left<#1\right>}

\newcommand{\ffrac}[2]{\frac{\displaystyle #1}{\displaystyle #2}}

\newcommand{\Np}{\ensuremath{N_{\mathrm{P}}}}        
\newcommand{\Nc}{\ensuremath{N_{\mathrm{C}}}}        
\newcommand{\N}{\ensuremath{N}}                                
\newcommand{\pp}{\ensuremath{\alpha_{\mathrm{P}}}}   
\newcommand{\pc}{\ensuremath{\alpha_{\mathrm{C}}}}   
\newcommand{\vp}{\ensuremath{v_{\mathrm{P}}}}          
\newcommand{\vc}{\ensuremath{v_{\mathrm{C}}}}          

\begin{document}
\title{
Spatially Resolving the Condensing Effect of Cholesterol in Lipid Bilayers
}
\author{Felix Leeb}
\author{Lutz Maibaum}
\email{maibaum@uw.edu}
\affiliation{Department of Chemistry, University of Washington, Seattle, WA 98195 }
\begin{abstract}
We study the effect of cholesterol on the structure of dipalmitoylphosphatidylcholine (DPPC) phospholipid bilayers. Using extensive molecular dynamics computer simulations at atomistic resolution we observe and quantify several structural changes upon increasing cholesterol content that are collectively known as the condensing effect: a thickening of the bilayer, an increase in lipid tail order, and a decrease in lateral area. We also observe a change in leaflet interdigitation and a lack thereof in the distributions of DPPC head group orientations. These results, obtained over a wide range of cholesterol mole fractions, are then used to calibrate the analysis of phospholipid properties in bilayers containing a single cholesterol molecule per leaflet, which we perform in a spatially resolved way. We find that a single cholesterol molecule affects phospholipids in its first and second solvation shells, which puts the range of this interaction to be on the order of one to two nanometers. We also observe a tendency of phospholipids to orient their polar head groups toward the cholesterol, which provides additional support for the umbrella model of bilayer organization. 
\end{abstract}

\maketitle

\section{\label{sec:intro}Introduction}

Lipid bilayers are fascinating materials due to their unusual physical properties and to their role as the foundation of biological membranes. While the behavior of pure bilayers is relatively well understood, that of mixed bilayers continues to be an important topic of ongoing research. Cellular membranes can contain dozens of different types of lipids, which give rise to complex spatial organization and lipid-specific effects on membrane proteins. This chemical diversity, however, makes it difficult to systematically investigate the physical effects that emerge when different types of lipids are mixed in a single bilayer. The study of model membrane systems that contain only a small number of lipid types has therefore been a fruitful approach to investigate mixed bilayers.

Of fundamental interest are binary lipid bilayers that contain a single type of phospholipid and cholesterol. It has been known for nearly a century that cholesterol significantly changes the properties of lipid films~\cite{Leathes25}. A prominent example is the condensing effect, which describes the decrease in lateral phospholipid cross sectional area induced by cholesterol. This is accompanied by an increase in ordering of the lipids' alkyl tails and a concomitant increase in lipid height. The condensing effect has been measured in multiple experimental~\cite{McIntosh78,Pencer05,Hung07, Kucerka08,Hung16} and computer simulation\cite{Smondyrev99,Rog01,Chiu02,Hofsaess03,Falck04,Edholm05,Berkowitz09,Bennett09c,Rog09,Alwarawrah10} studies. This body of work provides a detailed perspective on the effects of cholesterol on average membrane properties.

In this contribution we spatially resolve the ordering effect of a single cholesterol molecule on lipids in a dipalmitoylphosphatidylcholine (DPPC) bilayer. We do so by analyzing extensive all-atom molecular dynamics (MD) computer simulations. By considering the infinite dilution limit we can unambiguously determine the range of this peculiar interaction, and the extent to which a single cholesterol molecule can alter the structural properties of nearby phospholipids.

Our approach also provides novel opportunities to test phenomenological models of cholesterol--phospholipid interactions. For example, the umbrella model argues that cholesterol molecules minimize unfavorable solvent interactions by associating with phospholipids whose large polar head groups contribute to the shielding of the cholesterol's hydrophobic core~\cite{Huang99}. At finite concentration multiple cholesterol molecules compete for the limited space under a phospholipid's protective umbrella, which leads to additional effects such as frustrated interactions. By focusing on bilayers containing only a single cholesterol molecule per leaflet we obtain an unobstructed view of cholesterol's ordering effect on nearby phospholipids.

Before discussing the results of our spatially-resolved analysis of single-cholesterol simulations, we first present  results for the average structural properties of DPPC bilayers that contain from 0 to 50\% cholesterol. We will later use these results as a basis for comparison of the single-cholesterol simulations. These results are generally in agreement with previously reported studies, and we contribute novel results for the interdigitation of cholesterol-containing DPPC bilayers.

\section{\label{sec:methods}Methods}

\subsection{Computer Simulations}

Initial configurations for MD simulations were constructed using the Membrane Builder module of the Charmm-Gui molecular modeling server, Version 1.7~\cite{Wu14, Lee16}. Each system consisted of a symmetric lipid bilayer with $\Np$ DPPC and $\Nc$ cholesterol molecules per leaflet, resulting in $\N = \Np + \Nc = 50$ lipids per leaflet. The bilayer was solvated by 3200 water molecules. Lipids and water molecules where modeled by the Charmm36 and TIP3P force fields, respectively~\cite{Klauda10}.

All simulations were performed using the Gromacs 5.1 molecular dynamics simulation software~\cite{Abraham15} with a time step of 2 femtoseconds. Initial energy minimization and equilibration were performed according to Membrane Builder's default settings. A system temperature of $T = 323.15\,\mathrm{K}$ was maintained by a Nose-Hoover thermostat that coupled separately to the bilayer and solvent subsystems with a relaxation time of 1 ps~\cite{Nose84,Hoover85}. A Parrinello-Rahman barostat with semi-isotropic coupling (time constant 5 ps, compressibility $4.5 \times 10^{-5} \, \text{bar}^{-1}$) maintained an overall pressure of 1 bar and a vanishing surface tension~\cite{Parrinello81}. Dispersion forces were smoothly attenuated over a distance from 1.0 to 1.2 nm. Electrostatic interactions were computed using the Particle Mesh Ewald method~\cite{Darden93,Essmann95}. Covalent bonds that include a hydrogen atom were constrained using the LINCS algorithm~\cite{Hess97}.

The cholesterol mole fraction $\chi = \Nc / \N$ of the studied bilayers ranges from 0 to 50\%. The lengths of the simulated trajectories are \SI{400}{ns} for $\Nc\!=\!0$, \SI{3}{\micro \second} for $\Nc\!=\!1$,  \SI{1.5}{\micro \second} for $\Nc\!=\!2$, and \SI{1}{\micro \second} each for $\Nc\!=\!3, 5, 7, 10, 15, 20, 25$.

\subsection{Data Analysis}

For each simulated trajectory several globally averaged bilayer properties were computed. Bilayer thickness was calculated as the combined distance of phosphorus atoms of lipids from the two leaflets from a configuration's bilayer mid-plane. The extent of interdigitation of the two leaflets was computed as~\cite{Olsen13}
\begin{equation}
I = \ffrac{2 \int \dif z \, \min \left\{\rho_1(z),\rho_2(z)\right\}}{\int \dif z \, \rho_1(z) + \int \dif z \, \rho_2(z)} \label{eq:interdigitation}
\end{equation}
where $\rho_1(z), \rho_2(z)$ are the mass density profiles of the phospholipids in the upper and lower leaflet, respectively.

The ordering of the phospholipid tails was measured by calculating the order parameter
\begin{equation}
S_{\text{CH}}=\frac{1}{2}\left(3\left<\cos^2\theta_{\text{CH}}\right>-1\right) \label{eqn:tail}
\end{equation}
for every carbon atom in both acyl chains. Here $\theta_{\text{CH}}$ is the angle between the carbon-hydrogen bond vector and the outward vector normal to the bilayer, which we take to be parallel to the coordinate system's $z$-axis. This order parameter is the equivalent of the one frequently measured in deuterium NMR experiments~\cite{Vermeer07}.

To quantify the surface area per lipid we use the two-dimensional Voronoi tessellation of each leaflet with the same procedure as Pandit and coworkers~\cite{Pandit04}. This tessellation is constructed from a chosen set of vertices. It partitions the total area into segments, each of which is assigned to a specific vertex. To take into account the size difference between the phospholipid and cholesterol we use three vertices for DPPC (the middle carbon in the glycerol backbone and the first carbon in each tail) but only one vertex for the cholesterol (the oxygen atom). The areas of the three segments assigned to a single DPPC molecule were added to yield the phospholipid's Voronoi area $\vp$. These vertices were chosen because they lie roughly on the same plane at the interface between the hydrophobic and hydrophilic regions of the membrane~\cite{Pandit04}.

Uncertainties were estimated using the block average method described by Flyvbjerg and Petersen~\cite{Flyvbjerg89} and are often smaller than the point size used in the figures.

\section{\label{sec:averageproperties}Results: Average bilayer properties}

\subsection{Thickness and Interdigitation}

\begin{figure}
 \begin{center}
 \includegraphics[width=\columnwidth]{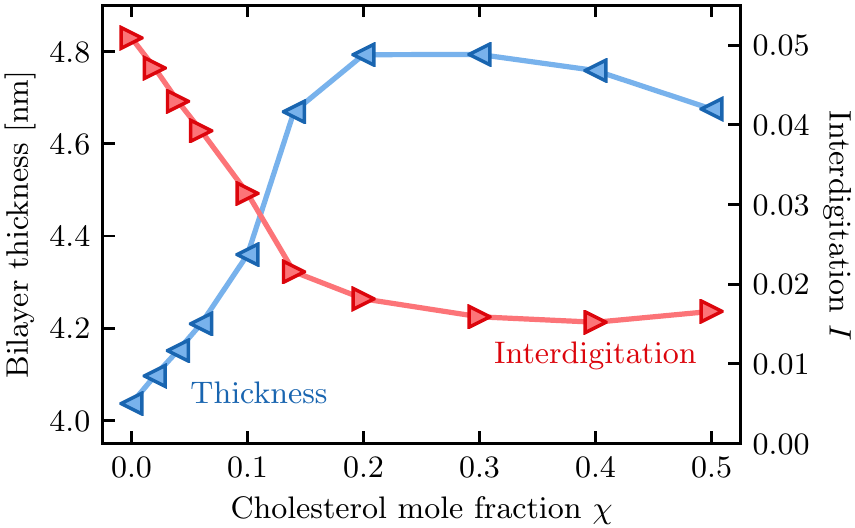}
\end{center}
\caption{\label{fig:thickness}
Dependence of bilayer thickness and leaflet interdigitation on composition in binary DPPC/cholesterol membranes. Starting from a pure phospholipid bilayer, thickness increases and interdigitation decreases significantly upon addition of cholesterol up to a mole fraction of 20\%. Both properties show a weak reversal of this trend at high cholesterol concentrations. 
}
\end{figure}

Figure~\ref{fig:thickness} shows how the thickness of the membrane, defined as the average distance between the phospholipids' phosphorus atoms of the two leaflets, varies with the cholesterol mole fraction $\chi$. As expected, we find that the bilayer thickness increases significantly as $\chi$ increases. We observe a thickening from \SI{4}{nm} for the pure phospholipid bilayer to \SI{4.8}{nm} at 20\% cholesterol content. At that concentration the thickness plateaus, and even slightly decreases at high cholesterol concentrations up to 50\%.

The increase in membrane thickness with cholesterol content is a well-known property of phospholipid bilayers. More surprising is the slight thinning at high cholesterol levels near $\chi = 0.5$. One can envision multiple potential mechanisms for this behavior. First, the two leaflets of the bilayer might interdigitate under these conditions. This behavior has been previously observed in computer simulations of unsaturated lipids~\cite{Olsen13} which exhibit a more pronounced thinning effect~\cite{Alwarawrah10}. Second, the decreased crowding in the hydrophilic region of the bilayer at large cholesterol mole fractions might allow the phospholipid head groups to bend towards the membrane plane to gain favorable interactions with the cholesterols' hydroxyl groups. Third, the ordering of the DPPC alkyl tails might partially reverse under those conditions, which would result in a decreased thickness of the hydrophobic part of the bilayer.

To test the first hypothesis we we calculated the extent $I$ of interdigitation, defined by \eqref{eq:interdigitation}. As shown in Figure~\ref{fig:thickness} we find that $I$ initially decreases with increasing cholesterol content, but stays essentially flat for $\chi \geq 30\%$. This suggests that increased interdigitation is not the reason for the slight decrease in bilayer thickness at high cholesterol concentrations. We will discuss the other two potential mechanisms in the following sections.

\subsection{Head Group Orientation}

\begin{figure}
 \begin{center}
 \includegraphics[width=\columnwidth]{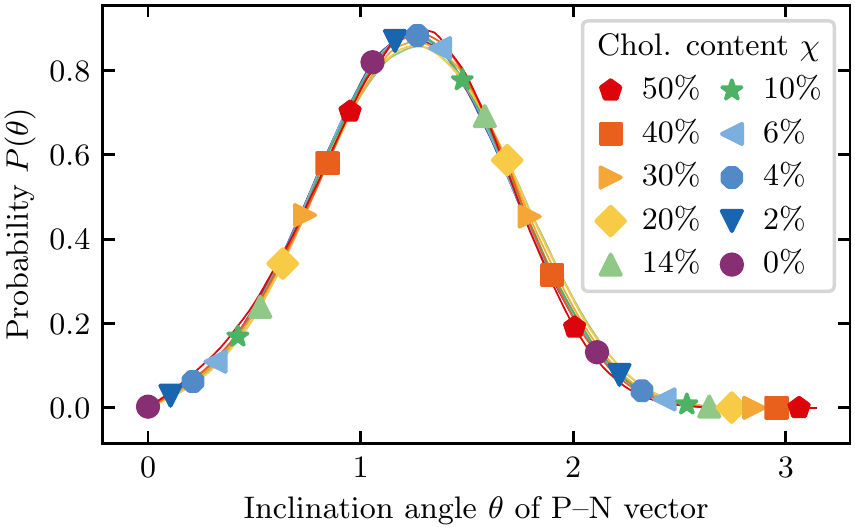}
\end{center}
\caption{\label{fig:PNangle}
Probability distributions of the inclination angle of the phospholipids' P--N separation vector for all 10 cholesterol mole fractions studied in this work. The distributions overlap and are nearly indistinguishable, showing that there are no significant changes in the head groups' tilt angles as the cholesterol content is varied. 
}
\end{figure}

To test whether changes in the phospholipids' head group tilt angle contributes to changes in bilayer thickness we calculate the probability distribution of the inclination $\theta$, defined as the angle between the vector from the phosphorus to the nitrogen atom and the outward membrane normal, which we assume to be parallel to the $z$ axis in our simulations. Our results are presented in Figure~\ref{fig:PNangle}, which shows that the distributions $P(\theta)$ remain unchanged for all ten bilayer compositions studied: the distributions are nearly indistinguishable from each other, and the average inclination angle varies only slightly from 70.4 to 72.8 degrees. Changes in head group inclination angles are therefore not responsible for the observed variation in bilayer thickness.

The invariance of head group orientations with cholesterol content seems to be inconsistent with the umbrella model, which posits that the phospholipids shield cholesterol molecules from contact with water by extending their large head groups over their smaller cohabitants. Our data shows that this effect, if it exists, does not affect the tilt angle distribution. However, as we will see in Section~\ref{subsec:LocalPNangle}, there are other aspects of head group orientation that indeed depend on the presence of cholesterol and that are consistent with the umbrella model.

\subsection{Lipid Tail Order}

\begin{figure}
 \begin{center}
\includegraphics[width=\columnwidth]{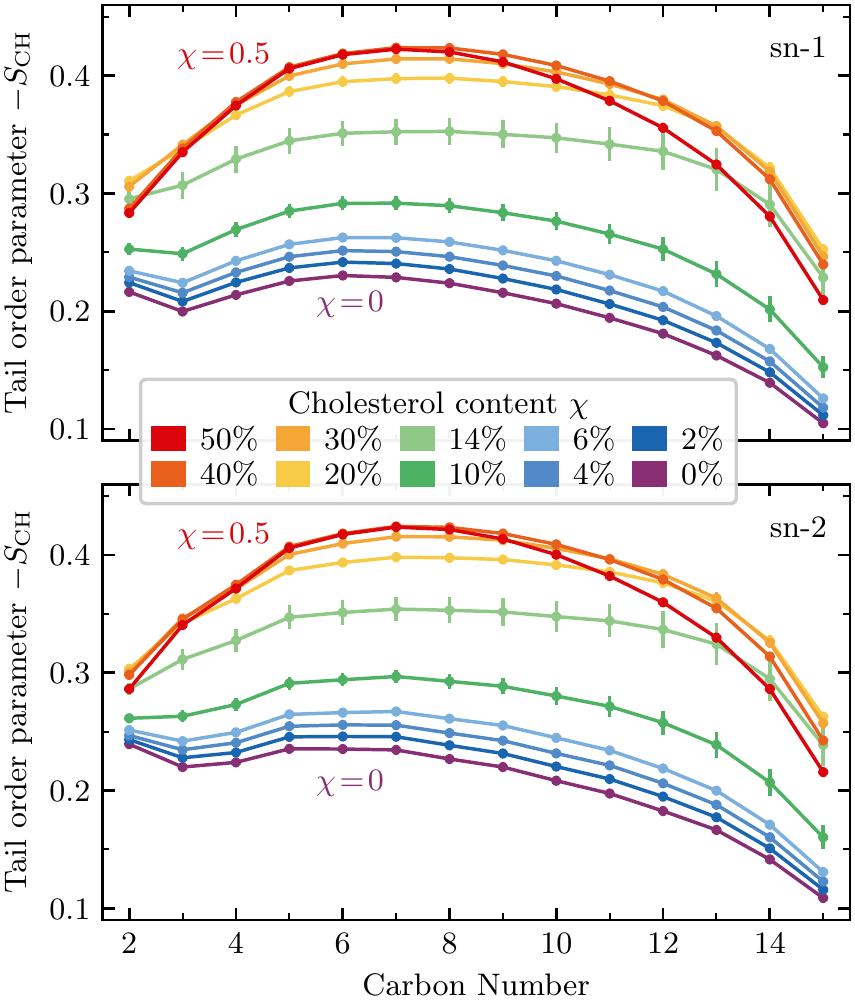}
\end{center}
\caption{\label{fig:order1}
Tail order parameters for the sn-1 (top panel) and sn-2 (bottom panel) chains of DPPC for all cholesterol concentrations studied. Starting from a pure DPPC bilayer, increasing the cholesterol content causes the entire tail to become significantly more ordered. At 20--30\% cholesterol content the ordering plateaus for the upper carbon atoms, but begins to decrease for carbon atoms near the bilayer center.
}
\end{figure}

To measure changes in the ordering of the phospholipid tails we compute the order parameter $S_{\text{CH}}$ for each carbon atom in DPPC's two alkyl chains. As shown in Figure~\ref{fig:order1}, we find that the entire hydrophobic tail initially becomes more ordered as the cholesterol content is increased, as expected from previous simulations and deuterium NMR experiments. At a cholesterol mole fraction of 20\% the ordering saturates for most carbons except for those near the center of the bilayer (positions 12 and below), which exhibit a decrease in order at high cholesterol levels.  Out of the three mechanisms considered, it is this reversal of ordering near the ends of the lipid tails that contributes the most to the apparent decrease in membrane thickness at high cholesterol concentrations.

The partial disordering of the phospholipid tails near the bilayer center can be understood by considering the relative lengths of cholesterol and DPPC molecules. As the height of the bilayer increases, the cholesterols' hydroxyl groups remains close to the phospholipids' head groups to form a polar surface that interacts favorably with the aqueous solution. Because cholesterol is shorter than a fully ordered DPPC lipid this generates additional space close to the bilayer center, which causes the observed disordering of the lipid tails' ends.

\subsection{Area per Lipid}

Perhaps the most striking effect that cholesterol has on phospholipid bilayers, and the one that gives the condensing effect its name, is the apparent decrease in molecular area as cholesterol is added to a pure lipid bilayer. Since its discovery~\cite{Leathes25} it has been measured by numerous methods both in experiments and in simulations, and several physical observables are frequently used in the literature.

\begin{figure}
 \begin{center}
\includegraphics[width=\columnwidth]{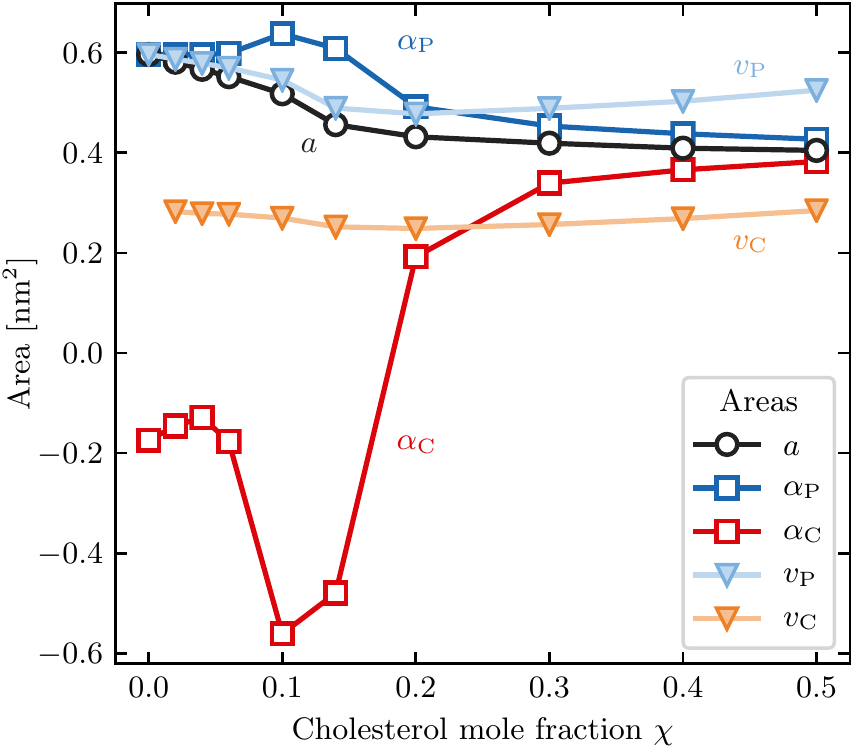}
\end{center}
\caption{\label{fig:averageareaperlipid}
Measures of area in binary DPPC/cholesterol bilayers. Shown are $a$, the average area per lipid (open circles), $\pp$ and $\pc$, the partial specific areas of phospholipids and cholesterol (open squares), and $\vp$ and $\vc$, the corresponding Voronoi areas (filled triangles). 
}
\end{figure}

The most apparent measure of the condensing effect is the area per lipid $a$, which is given by
\begin{equation}
a = \mean{A} / N
\end{equation}
where $\mean{A}$ the ensemble average of the bilayer area and $N = \Np + \Nc$ is the total number of lipid molecules (phospholipids and cholesterol) in each leaflet. This property depends on the composition of the bilayer, and Figure~\ref{fig:averageareaperlipid} shows this dependence of $a$ on $\chi$. We find that a pure DPPC bilayer has an area per lipid of \SI{0.60}{nm^2}, which is in good agreement with previous simulations utilizing the same force field, and only slightly below experimental measurements~\cite{Piggot12}. Raising the cholesterol content to 50\%, this area decreases monotonically, but not linearly, to \SI{0.41}{nm^2}.

As pointed out by Edholm and Nagle~\cite{Edholm05}, valuable information is contained in the nonlinearity of $a(\chi)$. Because the average bilayer area is an extensive function of the number of phospholipids and the number of cholesterol molecules, it follows from Euler's homogenous function theorem that it can be written as
\begin{equation}
\mean{A}  = \pp \Np + \pc \Nc \label{eq:Apartial}
\end{equation}
where
\begin{equation}
\pp = \ffrac{\partial \mean{A}}{\partial \Np}, \qquad \qquad \pc = \ffrac{\partial \mean{A}}{\partial \Nc}  \label{eq:partialspecificareas}
\end{equation}
are the partial specific areas of phospholipids and cholesterol, respectively. These properties depend on the composition of the bilayer, and one finds that~\cite{Edholm05}
\begin{eqnarray}
\pp(\chi) & = & a(\chi) - \chi \, a'(\chi) \\
\pc(\chi) & = & a(\chi) + (1-\chi) \, a'(\chi) 
\end{eqnarray}
where $a'(\chi) = \dif a / \dif \chi$ is the derivative of the average area per lipid with respect to cholesterol mole fraction.

Figure~\ref{fig:averageareaperlipid} shows our results for these quantities, where we have used a finite difference scheme to estimate the derivative $a'(\chi)$. It follows from the equations above that the partial specific area of DPPC is equal to the average area per molecule at $\chi=0$, and we find that this values decreases to \SI{0.43}{nm^2} at $\chi\!=\!0.5$. Cholesterol's partial specific area, on the other hand, has a negative value of \SI{-0.17}{nm^2} at $\chi\!=\!0$, indicating that adding a cholesterol molecule to a pure DPPC bilayer decreases its area. At higher cholesterol concentrations $\pc$ becomes positive, reaching \SI{0.38}{nm^2} at $\chi\!=\!0.5$.

At intermediate cholesterol concentration (10\% and 14\%) both $\pp$ and particularly $\pc$ show large fluctuations. We caution the reader that this is the region in which $a(\chi)$ has the largest curvature, and the estimator of the derivative might therefore be unreliable. A more accurate calculation would require additional simulations and more closely spaced composition values. Despite this caveat, it is clear that the partial specific area of cholesterol has the strongest dependence on composition, and that it changes sign from negative to positive over the range of cholesterol concentrations considered in our simulations.

An alternative approach to describe the effect of cholesterol on the area occupied by the different types of lipids is to partition the total bilayer area into segments that are considered occupied by a single molecule. This can be done, for example, using a Voronoi tessellation as described in the Methods section. In this case the area of a single simulation frame is equal to the sum of the areas of all the segments, and averaging over frames yields
\begin{equation}
\mean{A}  = \vp \Np + \vc \Nc
\end{equation}
where $\vp$ and $\vc$ are the segment areas averaged over all frames and all DPPC and cholesterol molecules, respectively.
This expression for the average bilayer area is superficially similar to equation~\eqref{eq:Apartial}, even though the partial specific areas and the Voronoi areas are substantially different quantities. The former are uniquely determined by the thermodynamic relationships~\eqref{eq:partialspecificareas} and can be negative, whereas the Voronoi areas are always positive by construction, and their values depend on the chosen method of area tessellation.

Our results for the Voronoi areas $\vp$ and $\vc$ are included in Figure~\ref{fig:averageareaperlipid}. While they change less dramatically with cholesterol content than their partial specific area counterparts, their variation suggests another trend: at low cholesterol levels, the average area of DPPC decreases more significantly than that of the cholesterol, which is explained by the straightening of the DPPC tails. However, at high cholesterol concentration, even though the overall area per lipid decreases as $\chi$ increases, both the average area per DPPC and the average area per cholesterol increase. This is consistent with the decrease in the bilayer thickness at high cholesterol concentrations, as it suggests that the tails of the phospholipids are maximally ordered between 20 to 30\% cholesterol. At even higher cholesterol mole fractions the DPPC tails become slightly less ordered, thereby increasing the area per lipid, but not by enough to mitigate the decrease in the total area of the leaflet due to the substitution of relatively large DPPC with smaller cholesterol molecules.

\section{\label{sec:resolvedproperties}Results: Local phospholipid properties}

Having established average bilayer properties for a wide range of cholesterol concentrations we now turn to the question to what extent a single cholesterol affects the phospholipids in its environment. We address this question by calculating some of the order parameter introduced in the previous section for subsets of DPPC molecules that are chosen by their distance from the cholesterol. This approach allows us to spatially resolve the condensing effect of a single cholesterol molecule.

We perform the analysis on the trajectory with $\Np=1$ cholesterol and $\Np=49$ phospholipid molecules per leaflet, and average all properties over both leaflets of the bilayer as before. While this bilayer has an average cholesterol mole fraction of 2\% it is not clear whether this small system is a good approximation of a macroscopic sample of the same average composition. If, for example, cholesterol were to cluster or interact in a significant way in the macroscopic system, its properties would deviate from those observed in our simulations where multiple cholesterols cannot interact with each other in the same leaflet. Our distance-dependent observations are therefore more akin to measurements in a macroscopic system in the infinite dilution limit (rather than at 2\% cholesterol mole fraction).

Because there is only a single cholesterol molecule in the simulated bilayer there are also very few phospholipids at any given separation from the cholesterol, and there is very little data available to average over. We address this difficulty by using a much longer trajectory of \SI{3}{\micro\second} than those used in the previous section for other bilayer compositions.

\begin{figure}
 \begin{center}
 \includegraphics[width=\columnwidth]{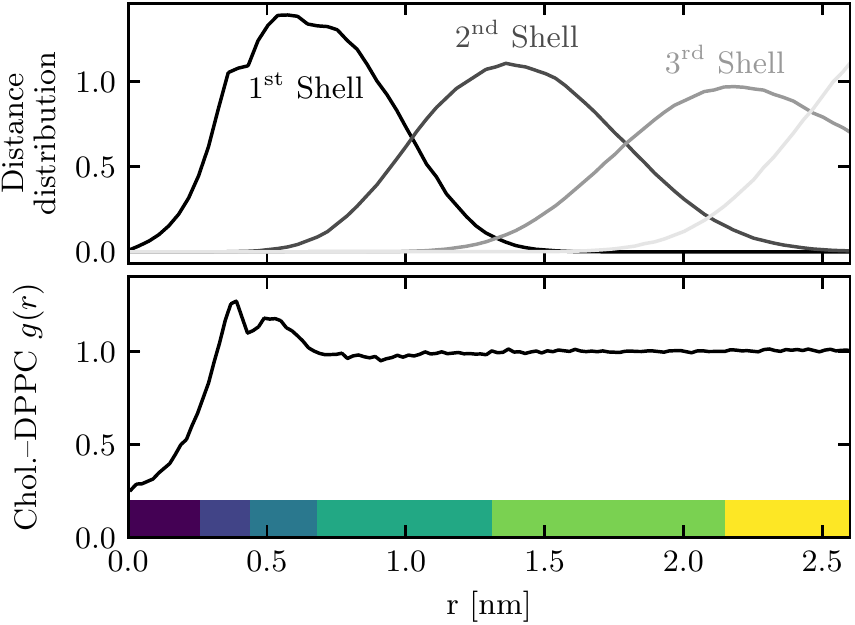}
\end{center}
\caption{\label{fig:gofr}
(Top panel) Distributions of the distances of DPPC phosphorus atoms from cholesterol's oxygen atom for each solvation shell as defined by the adjacency of the Voronoi tesselation.
(Bottom panel) Pair correlation function for these two atom types. The bottom patches show the six regions used in the analysis of local phospholipid properties in Section~\ref{sec:resolvedproperties}. 
}
\end{figure}

We begin our analysis by grouping DPPC molecules by their distance from the cholesterol molecule. The Voronoi tessellation provides a natural spatial ordering of lipids: those DPPC molecules whose Voronoi segments share an edge with that of cholesterol form the first solvation shell, those phospholipids that share an edge with the first shell form the second shell, and so forth. To give a sense of the sizes of these shells we show in Figure~\ref{fig:gofr} the distribution of the projected distance of DPPC phosphorus atom from the cholesterol oxygen atom for each solvation shell. We find that these distributions are rather broad, with the first solvation shell having significant contribution up to about \SI{1}{nm}.

As a separate measure of local structure we also compute the cholesterol--DPPC pair correlation function $g(r)$, also shown in Figure~\ref{fig:gofr}. This function shows a peak up to a separation of \SI{0.66}{nm} that indicates enhanced local structure, whereas $g(r)$ adopts is plateau value of unity at larger distances. At small distances the pair correlation function does not approach zero as it typically does for fluid systems, which is a testament to the fact that a two-dimensional description of bilayer systems is only approximate: a DPPC's phosphorus atom can indeed be located right on top of a cholesterol's oxygen when viewed in a two-dimensional projection.

\begin{figure}
 \begin{center}
 \includegraphics[width=\columnwidth]{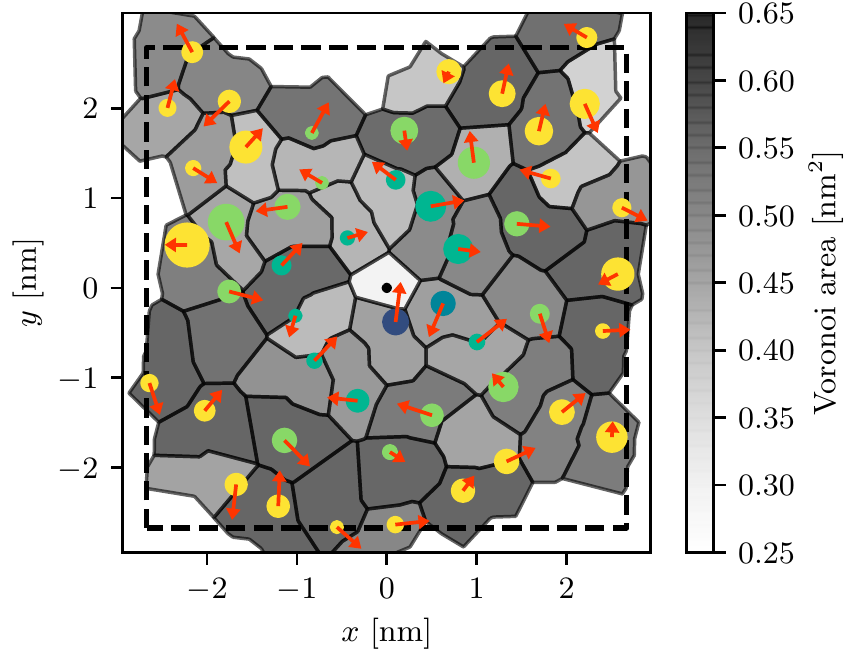}
\end{center}
\caption{\label{fig:vis}
Snapshot of one leaflet in a simulation with a single cholesterol molecule ($\Nc\!=\!1$) whose oxygen atom is at the center. Shown is the Voronoi tessellation of the bilayer, obtained by merging the three Voronoi segments of each phospholipid (see Methods). For each segment we show the position of the phosphorus atom by a disk whose size is proportional to the average tail order parameter $-S_{CH}$, and whose color corresponds to the six regions in the local analysis (Fig.~\ref{fig:gofr}). Also shown is the projection of the P--N separation vector for each phospholipid head group (arrow), and dashed lines are the boundaries of the periodically replicated simulation box. In this particular snapshot the head group of the DPPC molecule right below the central cholesterol is oriented toward the sterol.
}
\end{figure}

Based on this information we define six distinct regions of increasing distance from the cholesterol molecule, indicated in Figure~\ref{fig:gofr}. We divide the range of the peak in the pair correlation function into three regions, and group larger distances into another set of three regions. We then calculate for each lipid the area of its Voronoi segment, the tail order parameter, and the orientation of its head group as defined by the P--N separation vector. A simulation snapshot representing these properties is shown in Figure~\ref{fig:vis}. Structural properties are then averaged over lipids that are within the same distance region.

\subsection{Area per Lipid}

\begin{figure}
 \begin{center}
 \includegraphics[width=\columnwidth]{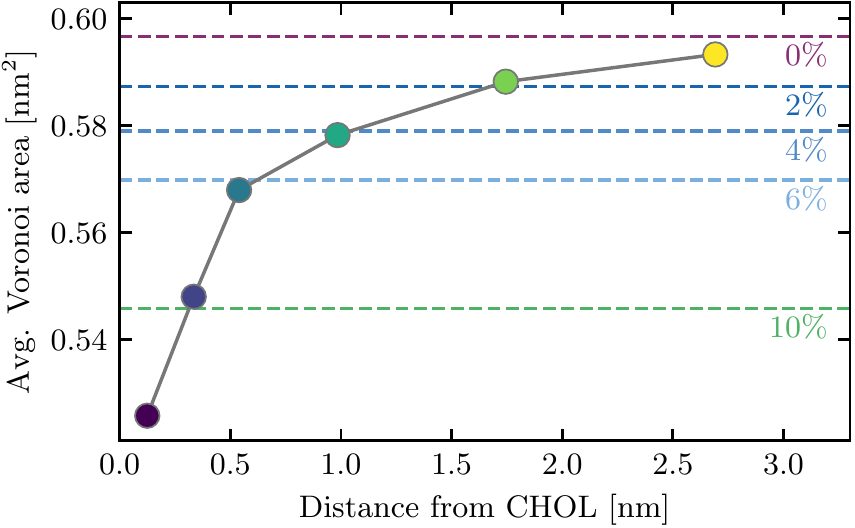}
\end{center}
\caption{\label{fig:localarea}
Average Voronoi area $\vp$ of phospholipids as a function of distance to the sole cholesterol molecule.  The initial sharp increase demonstrates the short range of the ordering effect. The dashed lines show the average Voronoi area of phospholipids in bilayers with the indicated cholesterol content for comparison.
}
\end{figure}

Figure~\ref{fig:localarea} shows the average Voronoi area of the phospholipids in each of the six regions of different separations from the cholesterol molecule. We find that a DPPC molecule in the first region occupies an average area of \SI{0.526}{nm^2}, whereas a phospholipid in the last region with an average distance of \SI{2.7}{nm} from the cholesterol has a Voronoi area of \SI{0.593}{nm^2}. The latter approaches \SI{0.60}{nm^2}, the average area we observed in a pure DPPC bilayer (Figure~\ref{fig:averageareaperlipid}). The former, on the other hand, is closer to the value of \SI{0.55}{nm^2} that we obtained for a bilayer with 10\% cholesterol.

This data illustrates that even a single cholesterol molecule has a significant condensing effect on the phospholipids in its immediate environment. A DPPC molecule in close proximity to the cholesterol occupies a similar, and even slightly smaller, area than it would typically in a bilayer with 10\% cholesterol content. This behavior can be rationalized by a simple argument: if a phospholipid in the first solvation shell has on average six nearest neighbors and one of those is a cholesterol molecule, its effective local composition is $1/6\!=\!16\%$ cholesterol.

The range of this local condensing effect is limited to one or two solvation shells of the cholesterol molecule; at distances exceeding \SI{2}{nm} the phospholipids' area is essentially indistinguishable from that in pure bilayer.

\subsection{Lipid Tail Order}

\begin{figure}
 \begin{center}
 \includegraphics[width=\columnwidth]{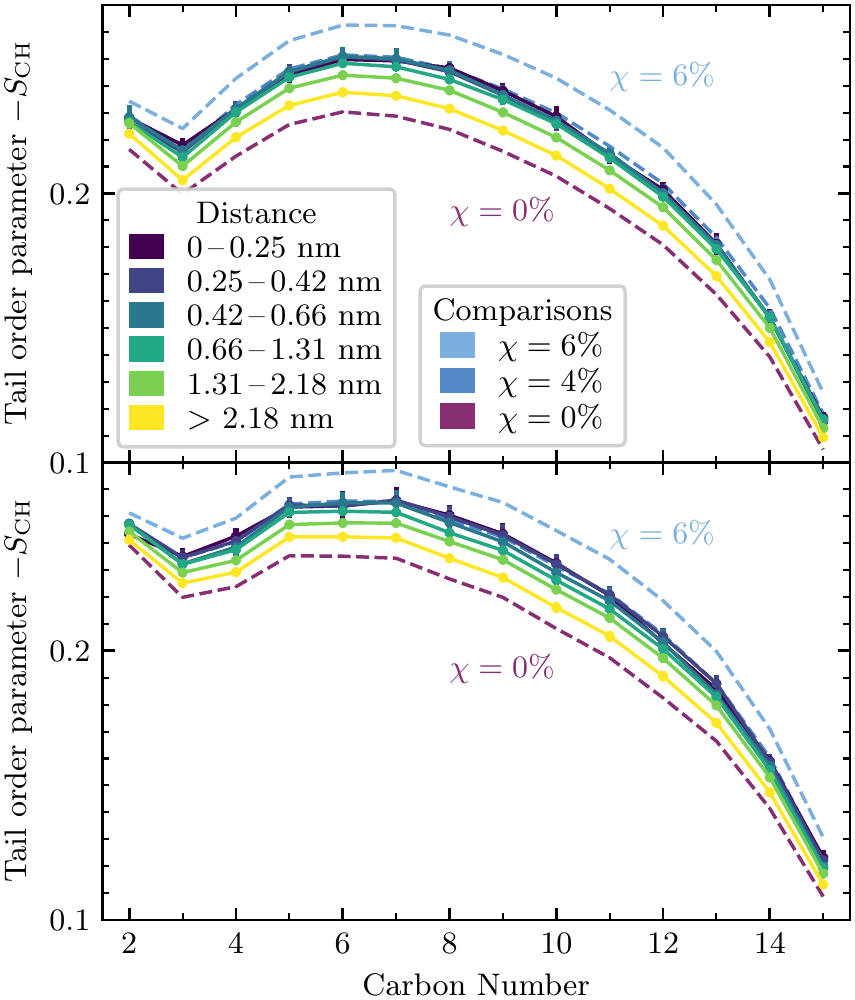}
\end{center}
\caption{\label{fig:localorder1}
Spatially resolved tail order parameters of the sn-1 (top) and sn-2 (bottom) alkyl chains of phospholipids near a single cholesterol molecule. Dashed lines show the average tail order parameter for bilayers containing, 0, 4, and 6\% cholesterol for comparison. Phospholipids in close proximity to a single sterol show slightly increased ordering, comparable to those in a bilayer with 4\% cholesterol.
}
\end{figure}

A similar but weaker condensing effect can be observed when measuring the ordering of the phospholipid tails in a distance-resolved manner. As shown in Figure~\ref{fig:localorder1}, DPPC molecules in the region closest to the cholesterol show a similar profile of the tail order parameter $S_\text{CH}$ as the average phospholipid in the 4\% cholesterol bilayer. Moving further away from the cholesterol the ordering of the tails decreases and approaches that of a pure phospholipid bilayer.

Given the simple argument put forth in the previous section it is surprising that the ordering of the phospholipid tails in the immediate environment of a cholesterol molecule is less pronounced than in a bilayer with a cholesterol content of 6, let alone 10 or 16, per cent. This suggests that unlike the Voronoi area, the ordering of the lipid tails depends more strongly on the global cholesterol concentration than on the local one, which suggests a cooperative nature of the lipid tail ordering.

\subsection{\label{subsec:LocalPNangle}Head group orientation}

\begin{figure}
 \begin{center}
 \includegraphics[width=\columnwidth]{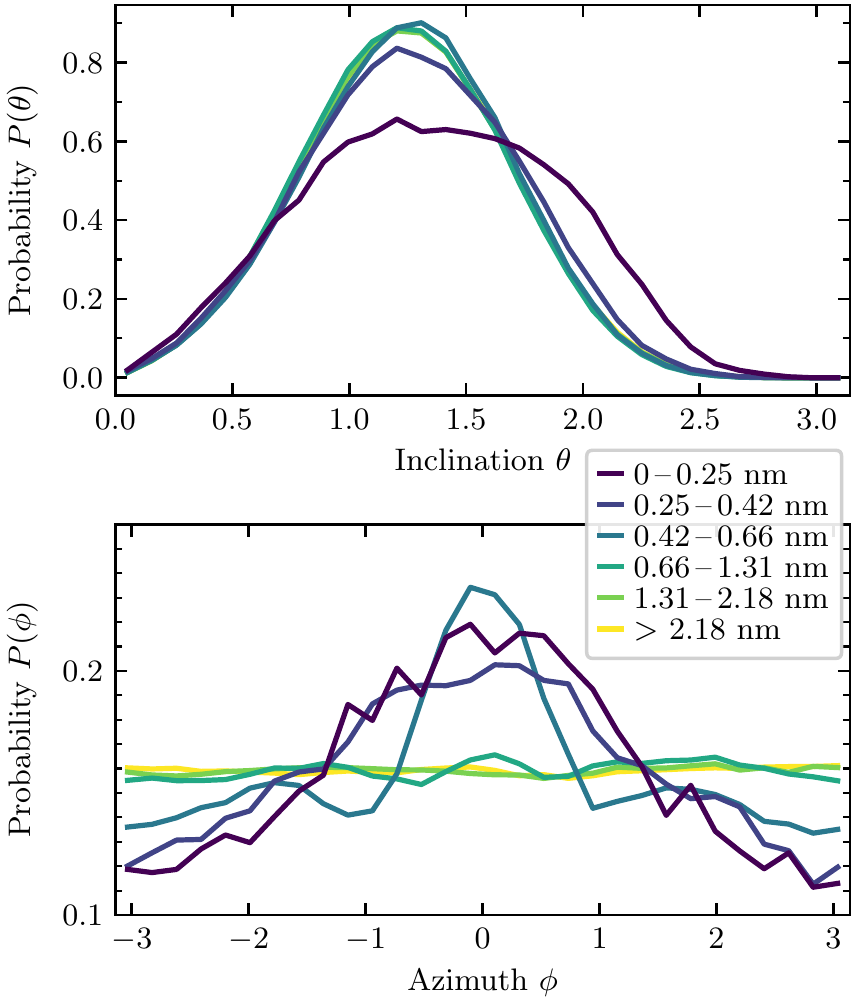}\llap{
 \raisebox{3.0in}{
   \includegraphics[width=0.82in]{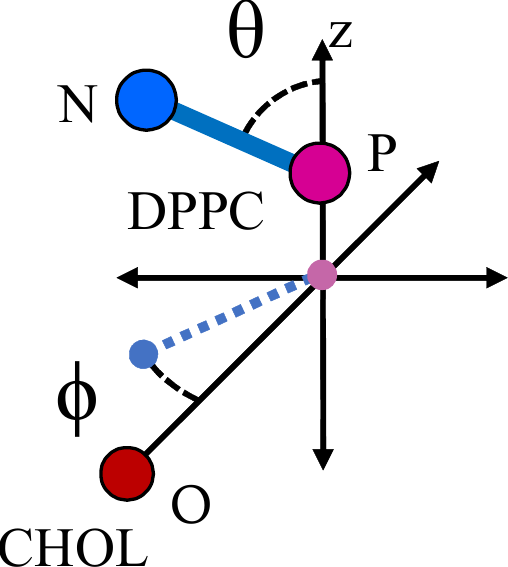}
   }
 }
\end{center}
\caption{\label{fig:phi}
(Top panel) Distribution of the DPPC head group inclination angle $\theta$ as a function of distance from a single cholesterol molecule. Except for phospholipids in the closest region, which show a broadening toward larger angles, the distributions remain similar to those obtained in pure bilayers. 
(Bottom panel) Distribution of the head group azimuth $\phi$, defines as the angle between the P--N separation vector of the head group and the O--P separation vector between cholesterol and DPPC, both projected onto the $xy$ plane (see inset).
Phospholipids within the first three regions, extending up to \SI{0.66}{nm}, have a tendency of orienting their head groups toward the cholesterol, whereas phospholipids further away show a uniform angle distribution.
}
\end{figure}

Finally we study how the presence of a single cholesterol molecule affects the orientation of the head groups of the phospholipids in its environment. As before (Figure~\ref{fig:PNangle}) we calculate the probability distribution of the inclination angle $\theta$ of the phosphate--nitrogen separation vector. However, due to the presence of a single cholesterol there now exists a unique reference point within the bilayer, and we can also measure the distribution of the azimuth angle $\phi$ of this vector. We choose the coordinate system such that a value of $\phi=0$ corresponds to a P--N vector that points toward the cholesterol molecule, as illustrated in Figure~\ref{fig:phi}.

We find that the distribution of the inclination is essentially the same as in a pure phospholipid for all but the closest region to the cholesterol molecule. In the latter case the distribution is shifted to slightly larger inclination angles, suggesting that the head groups are tilted more towards the bilayer plane.

More interesting is the distribution of the azimuth, shown in the bottom panel of Figure~\ref{fig:phi}. We find that phospholipids in the first three regions, extending to \SI{0.66}{nm} from the cholesterol, have a clear tendency to orient their polar head groups in the direction of the sterol. This result is a direct observation of the mechanism described by the umbrella model, which envisions phospholipid head groups to shield the sterol with its small polar group from unfavorable contacts with water~\cite{Huang99}.

As is expected, phospholipids farther away from the cholesterol show no preference for any particular orientation with respect to the cholesterol, as evidenced by completely uniform probability distributions.

\section{\label{sec:discussion}Discussion}

Our study of lipid structure in binary DPPC/cholesterol bilayers is consistent with most previous work on the condensing effect, and it highlights several new properties that emerge in these complicated systems. Our analysis of average lipid order recapitulates the well-known effects of adding cholesterol to phospholipid bilayers: the bilayer thickens and the average area per lipid decreases as the lipid tails become more ordered. In agreement with previous simulation studies we find that the partial specific area of cholesterol varies significantly, and even changes its sign, as the cholesterol mole fraction is increased from 0 to 50\%.

Less well understood is the importance of leaflet interdigitation, which we find to decrease up to a composition of 30\% cholesterol. Beyond that the extent of interdigitation remains approximately constant, in contrast to previous reports on unsaturated lipid bilayers. Whether this difference stems from the nature of the studied phospholipids or whether it is caused by differences in the underlying molecular mechanics force field remains to be explored. Within the set of our simulations, however, it is clear that increasing interdigitation is not the cause for the observed slight decrease in membrane thickness at high cholesterol levels, and neither is a changing distribution of head group inclination angles; the latter being essentially unaffected by the bilayer composition.

The analysis of simulation trajectories that contain only a single cholesterol per leaflet provides a novel way to characterize the condensing effect. We find that the average area of phospholipids in the first and, to a lesser extent, in the second solvation shell of a cholesterol are lower than they would be in a pure DPPC bilayer, and indeed are closer to the average area in a bilayer containing 10 to 14\% cholesterol. As shown in Figure~\ref{fig:localarea}, the range of induced order is limited to about \SI{2}{nm}, which sets the length scale of the condensing effect caused by a single cholesterol molecule. Beyond that distance the phospholipids' properties are the same as in a pure bilayer.

A similar but weaker ordering effect can be seen in the tail order parameter, measured as a function of distance from the cholesterol. We find that a phospholipid in close proximity displays an ordering that is typical for a bilayer containing 4\% cholesterol. The fact that such a DPPC molecule, which by selection is surrounded by exactly one cholesterol and phospholipids otherwise, shows significantly less tail ordering than a DPPC molecule that is surrounded by (on average) one cholesterol and otherwise by DPPC molecules that are themselves surrounded by one cholesterol (on average), shows that the tail ordering is a cooperative effect that not only depends on the distance to the cholesterol, but also on the properties of the surrounding phospholipids.

\begin{figure}
 \begin{center}
 \includegraphics[width=\columnwidth]{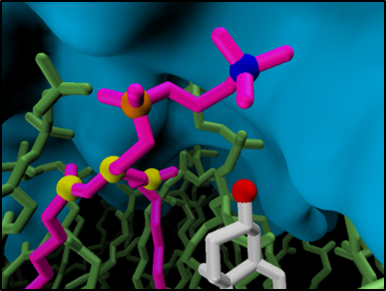}
\end{center}
\caption{\label{fig:snapshot}
Simulation snapshot of a DPPC bilayer containing a single cholesterol molecule, showcasing a phospholipid whose head group is oriented toward the cholesterol. It thereby shields the latter from contact with water, shown as the smooth isodensity surface in the background. Hydrogen atoms are omitted for clarity. Highlighted are the oxygen atom of cholesterol, the phosphate and nitrogen atom in the DPPC head group, and the three carbon atoms used in the Voronoi tessellation for the phospholipid.
} 
\end{figure}

The cholesterol also has a significant effect on the orientation of the head groups of nearby phospholipids. DPPC within about \SI{0.7}{nm} of the cholesterol show a strong preference for orienting the choline group towards the cholesterol. This is accomplished by a rotation around the membrane normal axis rather than an increase in inclination angle. A typical simulation snapshot of such a configuration is shown in Figure~\ref{fig:snapshot}. These results support the umbrella model of cholesterol--phospholipid interaction, which suggests hydrophobic shielding of the cholesterol by the DPPC head group as the driving force for the observed behavior.

Our approach of studying bilayers with a single cholesterol molecule allows us to isolate the direct ordering effect on a nearby phospholipid from collective effects caused by interactions with other phospholipids that are themselves in contact with cholesterol, and so forth. Because there are only very few (around 6) phospholipids in the first solvation shell of the sterol we need to perform rather long simulations to obtain a sufficient signal-to-noise ratio. To ease the computational burden we chose a small bilayer path of only 50 molecules per leaflet. Figures~\ref{fig:localarea}, \ref{fig:localorder1} and \ref{fig:phi} indicate that this system size is sufficient: the simulation box is large enough for the condensing effect to decay, and DPPC molecules that are more than \SI{2.2}{nm} separated from the cholesterol show nearly the same behavior as lipids in a pure bilayer.

As briefly mentioned in Section~\ref{sec:resolvedproperties}, a potential downside of this approach is that our simulations at very low cholesterol numbers might not be representative for larger systems with the same average cholesterol concentration. Several recent studies have highlighted the propensity of cholesterol to form transient clusters~\cite{Martinez10,Dai10,Bandara17}, which is suppressed in our calculations. As such our results should be viewed as observations of phospholipid behavior near isolated cholesterol molecules, which is realized in the low concentration limit of extended lipid bilayers.

\section*{Acknowledgments}

This work was facilitated though the use of advanced computational, storage, and networking infrastructure provided by the Hyak supercomputer system at the University of Washington.

\bibliography{references}

\end{document}